\documentclass[a4paper,10pt]{iopart}

\usepackage{iopams}

\expandafter\let\csname equation*\endcsname\relax

\expandafter\let\csname endequation*\endcsname\relax

\usepackage{amsmath}
\usepackage{amsmath}
\usepackage{latexsym}
\usepackage{amsfonts}

\usepackage{amssymb}
\usepackage{graphicx}

\usepackage[utf8]{inputenc}
\usepackage[T1]{fontenc}

\begin{document}

\title{Flux driven pedestal formation in tokamaks: turbulence simulations validated against the isotope effect}

\author{G. De Dominici$^1$, G. Fuhr$^2$, P. Beyer$^2$, C. Bourdelle$^1$, L. Ch{\^o}n{\'e}$^3$, F. Cianfrani$^2$, G. L. Falchetto$^1$, X. Garbet$^1$, Y. Sarazin$^1$}

\address{$^1$ CEA, IRFM, F-13108 Saint-Paul-lez-Durance, France}
\address{$^2$ CNRS, Aix-Marseille Univ., PIIM UMR7345, Marseille, France}
\address{$^3$ Department of Applied Physics, Aalto University, Espoo, Finland}

\begin{abstract}
The spontaneous pedestal formation above a power threshold at the edge of magnetically confined plasma is modelled for the first time in flux driven three-dimensional fluid simulations of electromagnetic turbulence with the code EMEDGE3D. The role of the collisional friction between trapped and passing particles is shown to be the key ingredient for shearing the radial electric field, hence stabilizing the turbulence, rather than the Maxwell and Reynolds stresses. The isotope effect, observed in many tokamaks worldwide, is recovered in EMEDGE3D simulations: the power threshold for  pedestal formation is lower for Tritium than for Deuterium. The  turbulence auto-correlation time is found to increase with the ion mass easing the radial electric shear stabilization, hence the pedestal formation.
\end{abstract}

\maketitle
\ioptwocol

Since its discovery in 1984 \cite{Wag84}, the H-mode has provided a route to substantially improve energy confinement in tokamak experiments \cite{Mar08,Car94,Beh14,Zha13} when the input power is higher than a certain threshold. The power thresholds obtained on various tokamaks have been gathered in an international database \cite{Tak04,Mar08}. The analysis of the latter has shown that higher thresholds are obtained with larger magnetic field, density, as well as with lower Hydrogenic isotope mass \cite{Ryt13,Mag18,Liu16}. Unfortunately, the scaling laws carry large uncertainties as the RMS value of usual fits is $\simeq30\%$ \cite{Mar08}.\\
Despite much progress in turbulence modelling \cite{Con00,Ras16,Cho14,Tam14,Par15,Cho15,Cha17,Gal17}, a first principle physics model of the L-H transition with predictive capability is still missing. Nonetheless, key players for realistic L mode edge turbulence drive have been identified \cite{ Rog98,Sco05} and the role of $\vec{E}\times\vec{B}$ shear stabilization leading to the formation of the edge transport barrier is long established both experimentally \cite{Bur97} and theoretically \cite{Big90}, for a review see \cite{Con00}.  Recently, both turbulence gyrofluid \cite{Ded19} and gyrokinetic models \cite{Bon19}, applied on JET and Asdex-Upgrade L mode edge parameters prior to a transition to H mode, have confirmed the predominance of resistive Drift Waves (DW) over Resistive Ballooning Modes (RBM), as well as of the destabilizing electromagnetic effects. Concerning the $\vec{E}\times\vec{B}$ shear, electrostatic flux-driven 3 dimensional simulations have shown that realistic edge collisional friction between trapped and passing particles is the key ingredient leading to $\vec{E}\times\vec{B}$ shear and hence triggering an edge transport barrier, provided that the input power exceeds some threshold \cite{Cho14,Par15,Cho15}.\\
In the present Letter, we show the first self-consistent pedestal formation in 3D non-linear fluid flux-driven simulation including the following critical physical ingredients: 1) RBM and DW 2) electromagnetic effects 3) a realistic neoclassical friction. As in experiments, the pedestal forms above a certain power threshold. As in experiments, this power threshold is lower for Tritium plasmas than for Deuterium plasmas.\\

\begin{figure}
\includegraphics[width=1.0\columnwidth]{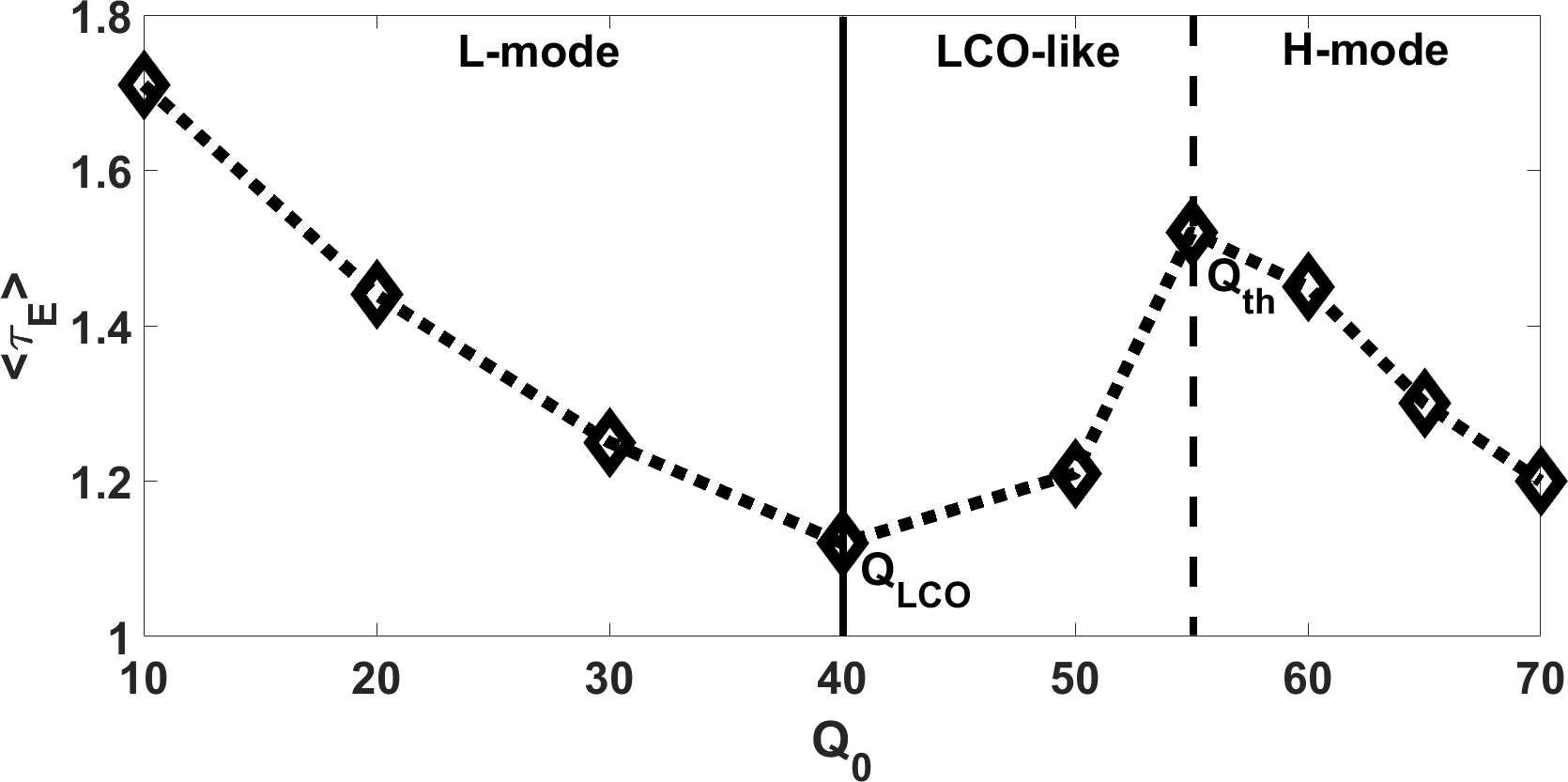}
\caption{\label{fig:Fig1} Time and radially averaged (over $0.86<r/a<1$) confinement time $\tau_E$ in ms with respect to the radially averaged heat flux $Q_0$ in arbitrary units.}
\end{figure}

\begin{figure}
\includegraphics[width=1.0\columnwidth]{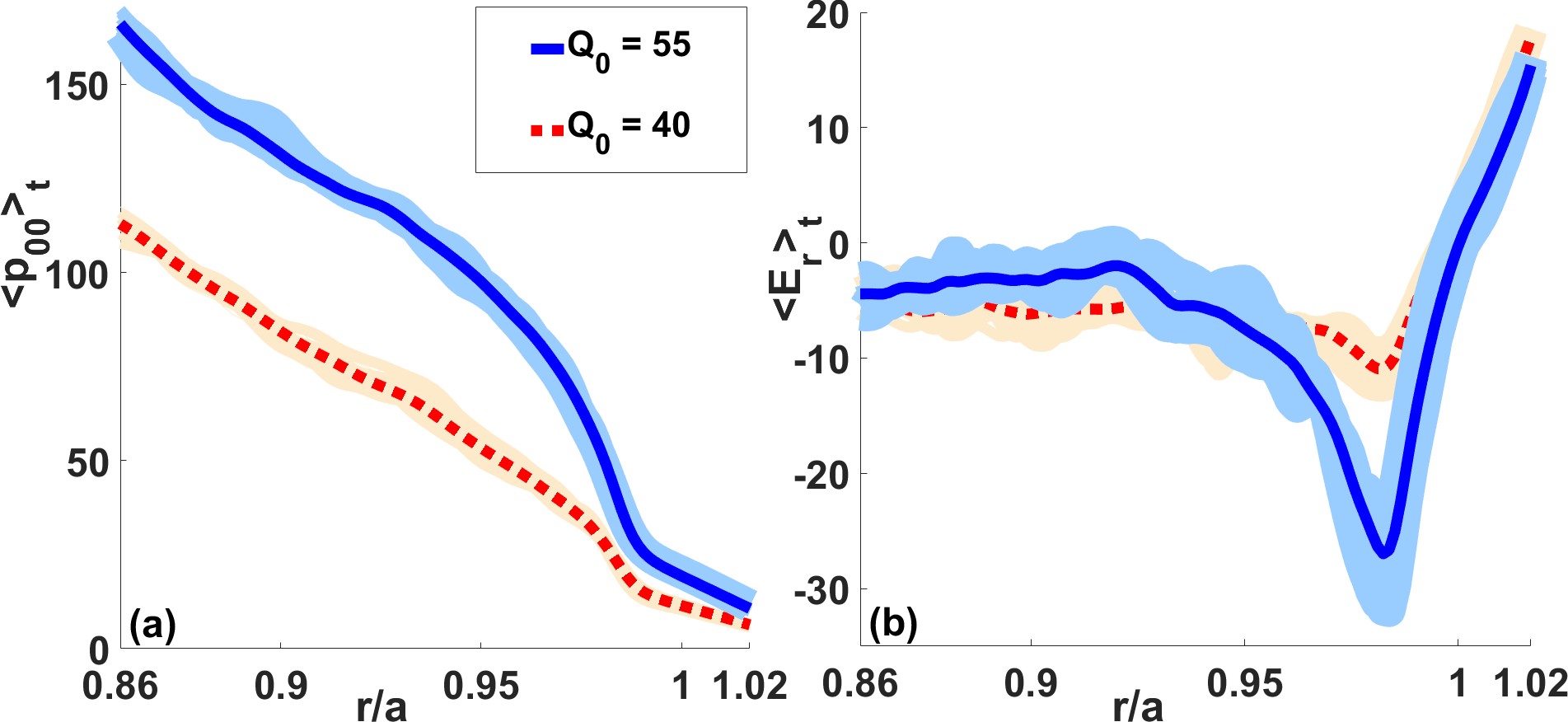}
\caption{\label{fig:Fig2} Time averaged profiles of, (a) the equilibrium pressure in arbitrary units, (b) the equilibrium radial electric field in arbitrary units, for a heat source lower than the threshold ($Q_0=40$, dashed line) and a heat source above the  threshold ($Q_0=55$, full line). $1<r/a<1.02$ corresponds to the RHS buffer zone.}
\end{figure}

The model of EMEDGE3D is based on nonlinear fluid equations for the charge, energy balance and Ohm's law, the three fields $\phi$, $p_e$ and $\psi$ being the electrostatic potential, the electron pressure and the magnetic potential \cite{Fuh08, Fuh08B} such that:
\begin{align}
\nonumber
\partial_{t}\nabla_\perp^2\phi+ \left\{\phi,\nabla_\perp^2\phi\right\} &=  - 
 {\nabla_\parallel}
\nabla_\perp^2\psi   - \omega_D  {\bf G}
p_e + \;\\[.25\baselineskip]
\label{eq:1}
&\mu_\perp \nabla_\perp^4\phi +\partial_r F_{neo}\;,\\[.25\baselineskip]
\nonumber
\partial_{t}p_e+\left\{\phi,p_e\right\} &= - \Gamma\nabla_{\parallel}\nabla_\perp^2\psi + \Gamma\omega_D {\bf G}\left( \phi - p_e\right)  +\;\\[.25\baselineskip]
\label{eq:2}
 &{\chi}_{\perp}\nabla_\perp^2 p_e + S\;,\\[.25\baselineskip]
\label{eq:3}
\partial_t \,\psi  &= -\tilde{\beta}_e^{-1} {\nabla_\parallel} (\phi-p_e) +\eta \nabla_\perp^2\psi \;
\end{align}
The toroidal magnetic equilibrium is 3D with circular and concentric closed magnetic flux surfaces, in the limit of large aspect ratio, in the slab approximation. The dimensionless coordinates $(x,y,z)$ refer respectively to the minor radius $r$, the poloidal and toroidal angles. $\nabla_\perp$ and $\nabla_\parallel$ are the perpendicular and parallel gradients with respect to the magnetic field lines and $G$ is the toroidal
curvature operator. $\{\phi,.\}$ is the Poisson bracket: $\{f,g\}=\partial_r f \partial_y g - \partial_y f \partial_r g$. $\mu_\perp$ is the classical viscosity, while $\chi_\perp$ accounts for perpendicular collisional heat diffusivity and $\eta$ for the resistivity. $\omega_D = 2L_\perp/L_\parallel$ corresponds to $2$ times the ratio between the typical perpendicular and parallel lengths with $L_\parallel=R_0$ (with $R_0$ the major radius) and $L_\perp=\rho_i$ (with $\rho_i$ the ion Larmor radius). Time, $t$, is normalized to the drift time $\frac{L_p}{c_s}$ with $c_s=\sqrt{\frac{T_e}{m_i}}$. $\Gamma=5/3$ is the heat capacity ratio and $\beta_e$ is the ratio of the electron kinetic pressure to the magnetic pressure. The density is kept constant $n=n_0$ in the simulation domain, and therefore $p=n_0T$ with $T=T_e=T_i$. The main simulation domain covers a radial extension between $0.86 \leq r/a \leq 1$ with $a$ the plasma minor
radius. It is bounded on both sides by buffer zones, where the diffusivity $\chi_\perp$ is artificially increased in order to prevent turbulence growth near the extremal points of the radial domain. The system is flux driven by a Gaussian shaped power source $S$ located in the inner buffer ($r/a < 0.86$) to simulate the heat deposition in the core plasma. This source imposes the influx $Q_0 = \int S(x)dx$ across the simulation domain, constraining the time averaged total heat flux at equilibrium to be equal to $Q_0$. The neoclassical friction term, $\partial_r F_{neo}$, ensures that the poloidal flow relaxes towards its neoclassical value on a collisional time; with $F_{neo}$, for a case without toroidal rotation, such that:
\begin{align}
\label{eq:4}
F_{neo} = -\mu_{neo}[\partial_r\bar{\phi}- K_{neo} \partial_r\bar{p}_e] 
\end{align}
Where $\mu_{neo} = \alpha_{neo}\mu_i[q(r)/\epsilon(r)]^2$ with $q$ the safety factor, $\epsilon$ the inverse aspect ratio and $\alpha_{neo}$ expected to be equal to unity but kept as a free parameter in the model, the neoclassical viscosity, $\mu_i$, is based on the heuristic closure proposed in \cite{Gia02} and $K_{neo}=1-k_i$ with $k_i$ a smooth fit between asymptotic collisional regimes from \cite{Hin76}.\\
For all simulations presented in this section, 190 points in the radial domain are used, with $\Delta x = 0.58$. The box
size is $L_r = 0.1m$. The spectral domain in the poloidal and toroidal directions is defined by $(m_{max}, n_{max}) = (250, 72)$ with $\Delta n = 4$. The dimensionless parameters used are based on a JET L mode edge prior to a transition in H mode \cite{Bou14,Bou15}, except for the temperature which is $\simeq 2$ times lower and the effective charge which is here 1, hence, overall, a realistic resistivity is used: $T_{i} = T_{e} = 50 eV$ , $n = 2.5 \times 10^{19} m^{-3}$, $B_0 = 1 T$, $R_0/L_p = 58$, $\beta_e = 0.2\%$, $\eta_0 =  2\times 10^{-6} \Omega.m$, $R_0 = 1.75 m$ and $a=0.45 m$, the safety factor q varies between $2.5$ and $3.5$ in the main simulation domain with a hyperbolic profile along the radial direction. The normalized poloidal wave vector $k_y=\frac{m\rho_i}{r_0}$ varies in  $[0,0.5]$. Averages are performed over $1000$ time units. The ion mass is taken to $2m_p$ (with $m_p$ the mass of a proton) for deuterium, and $3m_p$ for tritium.\\
We start using Deuterium plasmas, the initial state is a random phase distribution for the fluctuating fields and the heat flux $Q_0$ is scanned. The confinement time is defined within the simulated volume as: $\tau_E=\int_{r_{min}}^{r_{max}}P_{00}(r)dr/Q_0$ where $P_{00}(r)$ is the equilibrium pressure profile. As the other plotted quantities, $\tau_E$ is also averaged over 1000 time units. Nine nonlinear flux driven simulations are performed for the normalized flux, $Q_0$, varying from 10 up to 70. On Fig.~\ref{fig:Fig1}, one can see first the decrease of the confinement time as the heat flux increases, as expected from scaling laws \cite{iter-06}. When $Q_0$ exceeds $40$, a sharp increase of the confinement time by $40\%$ is observed. Above $Q_0=55$, the expected $\tau_E$ decrease with increasing $Q_0$ is recovered. The increase of confinement above $Q_0=40$ is due to the formation of a steep edge pressure gradient, as illustrated on Fig.~\ref{fig:Fig2}. The steepening of the pressure gradient at the edge is called a pedestal and is characteristic of the L-H transition. For profiles at $40\leq Q_0<55$ the pedestal exhibits quasi-periodic relaxations on a time scale smaller than $0.1$ms, reminiscent of Limit Cycle Oscillation (LCO) \cite{Sch14}. 


\begin{figure}
\includegraphics[width=1.0\columnwidth]{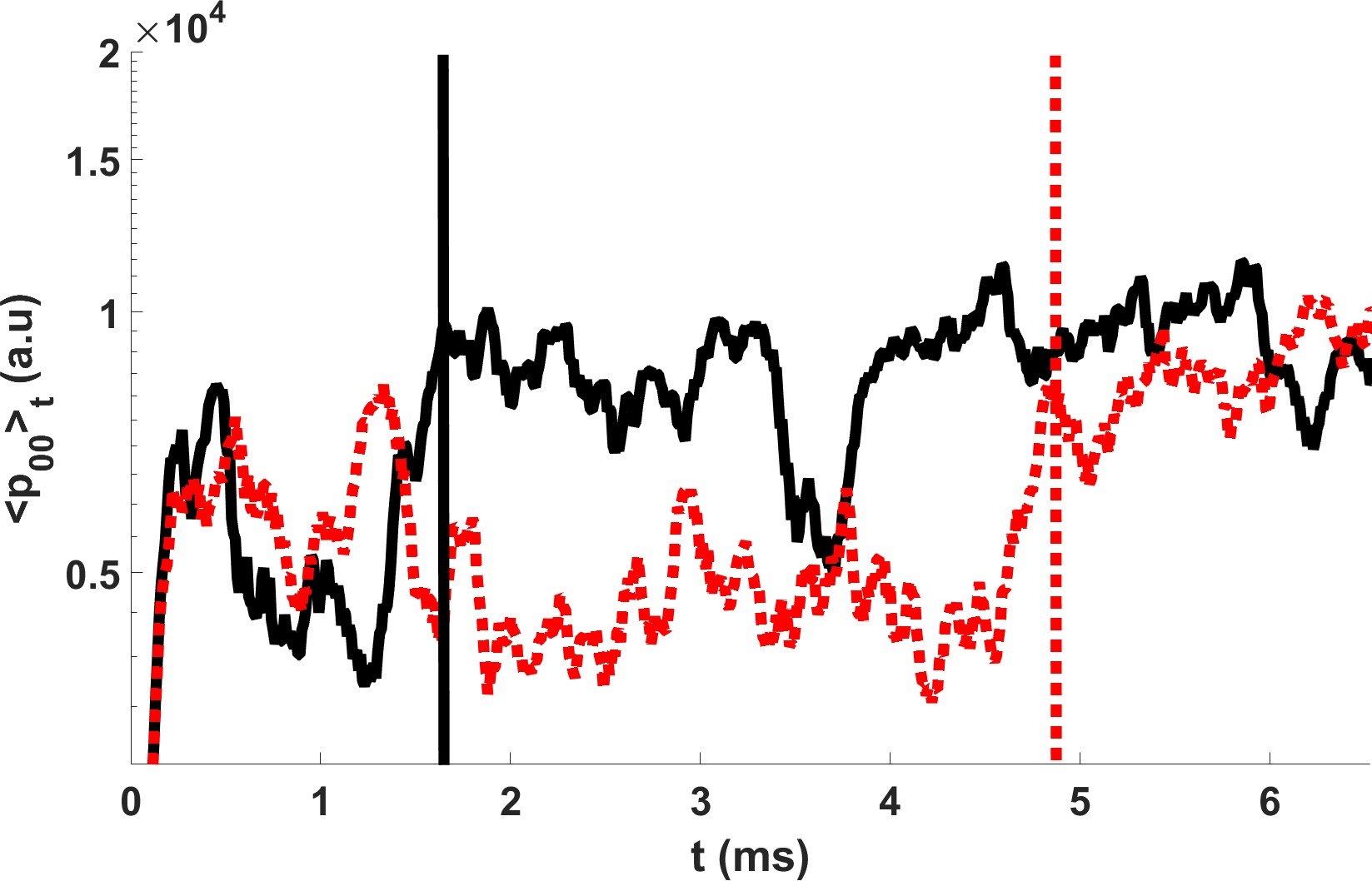}
\caption{\label{fig:Fig3} Energy of the equilibrium pressure as a function of time, $<P_{00}>_t=\int_{r_{in}}^{r_{out}}P_{00}(r)dr$ for  $\alpha_{neo}= 0.9$ (dashed line) and $2.8$ (full line) and $Q_0=70$ for both cases.}
\end{figure}

In Fig.~\ref{fig:Fig2}, the pressure profiles at $Q_0=40$ and $Q_0=55$ are compared. Inside $r/a<0.95$, the normalized gradient does not vary radially and only marginally with $Q_0$. For $Q_0 = 55$, a pedestal forms in the very edge region, and the mean radial electric field profile becomes hollow at $0.95 < r/a < 1$ generating a stabilizing localized shear flow \cite{Hin93, Bey05, Str13}. The shape of the $E_r$ profile shows good qualitative agreement with measurements in H-mode \cite{Sau12, Vie13} (see Fig.~\ref{fig:Fig2}b).  Note that $E_r$ is allowed to vary freely inside the buffer zone and exhibits a positive value in the outer buffer region. Note that a positive $E_r$ at the LCFS is expected Scrape-Off Layer physics \cite{Loi13}.\\
 The free parameter of the model, $\alpha_{neo}$, determines how fast $E_r$ relaxes to its neoclassical equilibrium value. It is therefore important to check wether the choice for $\alpha_{neo}$ impacts the energy level on which the simulation converges. Fig.~\ref{fig:Fig3} shows the energy of the equilibrium pressure as a function of time for $Q_0=70$ and two values of $\alpha_{neo}$, 0.9 and 2.8. For $\alpha_{neo}= 0.9$, the pedestal formation occurs at t = 4.8ms while it occurs sooner, at 1.7ms, for $\alpha_{neo}=2.8$. Once the pedestal appears, both cases exhibit the same energy level and the same pressure profile. In order to save CPU time, $\alpha_{neo}=2.8$ is used throughout the paper.\\

To compare the contributions of the Reynolds Stress (RS), Maxwell Stress (MS) and neoclassical friction before and after the pedestal formation, we derive  an  equation  governing the  $\vec{E}\times\vec{B}$ flow \cite{Cho15},  by taking the flux surface average of equation~(\ref{eq:1}) and integrating along x:

\begin{align}
\partial_t\bar{u}_{Ey} = &-\partial_r\left<\tilde{u}_{Ey}\tilde{u}_{Ex}\right>_{y,z}\nonumber \\
&+ \hat{\beta}_e\partial_r\left<\tilde{b}_{y}\tilde{b}_{x}\right>_{y,z}\nonumber \\
&   + \omega_d \partial_r\left<\tilde{p}_e\sin(k_y y)\right>_{y,z}\label{eq:5}\\
& + F_{neo}\nonumber \\
&+ \nu_{\perp}\partial_{r}^2 \bar{u}_{Ey}\nonumber
\end{align}

\begin{figure}
\includegraphics[width=1.0\columnwidth]{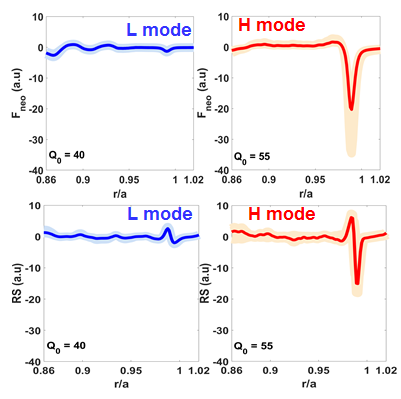}
\caption{\label{fig:Fig4} Time averaged profiles of contributions to the  $\vec{E}\times\vec{B}$ shear flow production, see equation~\ref{eq:5}. Upper panels: the neoclassical friction contributions. Lower panels: the Reynolds Stress contributions. Left: before the transition, $Q_0=40$, right: after the transition, $Q_0=44$.}
\end{figure}

The terms on the RHS of equation~\ref{eq:5} are the sources and sinks of the flow. The first RHS term relates to the contribution from RS, the second from the MS, the third from the curvature of the B field, the fourth from the neoclassical friction and the fifth from the viscosity. In the simulations,  below the transition as well as above, the MS and the curvature contributions are both negligible compared to the RS and the neoclassical friction contributions. Moreover, the RS and the neoclassical friction contributions are of similar amplitude as illustrated on Fig.~\ref{fig:Fig4}, both before and after the transition. It follows that the neoclassical friction acts as a key player in the pedestal formation. The results obtained in electrostatic simulations \cite{Cho14,Cho15,Par15} are here confirmed in the electromagnetic framework.\\

\begin{figure}
\includegraphics[width=1.0\columnwidth]{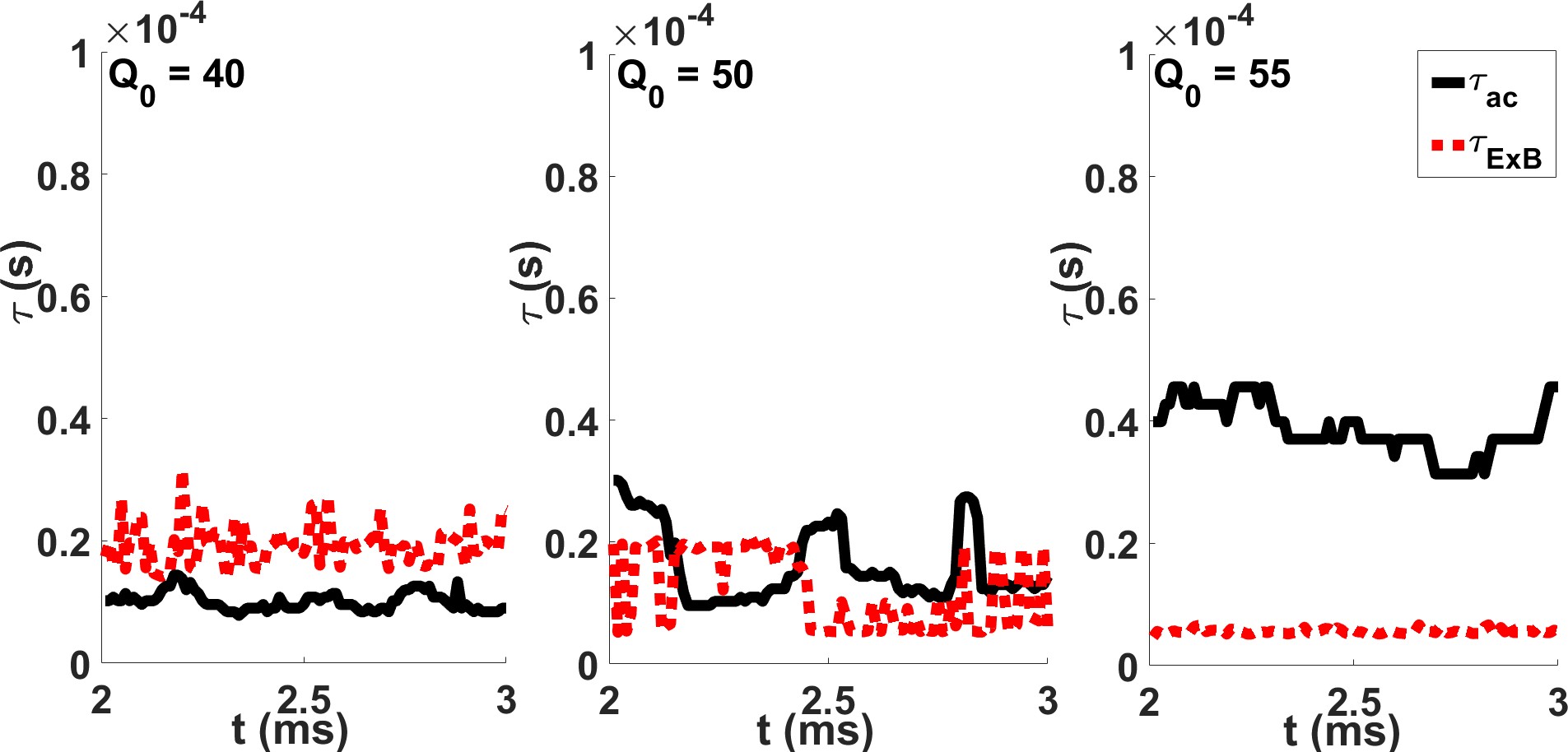}
\caption{\label{fig:Fig5} The turbulence auto-correlation time, $\tau_{ac}$ (full lines), the $\tau_{\vec{E}\times\vec{B}}$ the $\vec{E}\times\vec{B}$ shearing time (dashed lines) averaged over $0.95<r/a<1$, for three values of the heat flux $Q_0=40$ (left), $Q_0=50$ (center) and $Q_0=55$ (right).}
\end{figure}

The  physics  behind  the $\vec{E}\times\vec{B}$ shear stabilization is well captured by the BDT criterion  \cite{Big90} leading to turbulence suppression if:

\begin{align}
\label{eq:6}
\tau_{\vec{E}\times\vec{B}}<\tau_{ac}
\end{align}
With $\tau_{ac}$ the turbulence auto-correlation time and $\tau_{\vec{E}\times\vec{B}}\simeq\frac{1}{{V_{\vec{E}\times\vec{B}}}\prime}$, with  ${V_{\vec{E}\times\vec{B}}}\prime$ the  mean radial  derivative  of  the  $\vec{E}\times\vec{B}$ velocity in the pedestal region $0.95<r/a<1$ taken every 10 time units. The same average is performed on $\tau_{ac}$ which is the  e-folding  time  of  the auto-correlation function $C(d_t)=\frac{<\Phi(r,y,z,t)\Phi(r,y,z,t+dt)>}{<\lvert\Phi(r,y,z,t)\rvert^2>}$. On Fig.~\ref{fig:Fig5}, $\tau_{ac}$ and $\tau_{\vec{E}\times\vec{B}}$ are plotted for three values of increasing $Q_0$, in L mode, prior to the pedestal formation at $Q_0=40$, in the LCO-like phase at $Q_0=50$ and once the pedestal is formed for $Q_0=55$. One can see that the BDT criterion is indeed fulfilled only once the pedestal is formed, confirming the key role of the $\vec{E}\times\vec{B}$ velocity shear on the turbulence stabilization leading to the pedestal formation.\\

\begin{figure}
\includegraphics[width=1.0\columnwidth]{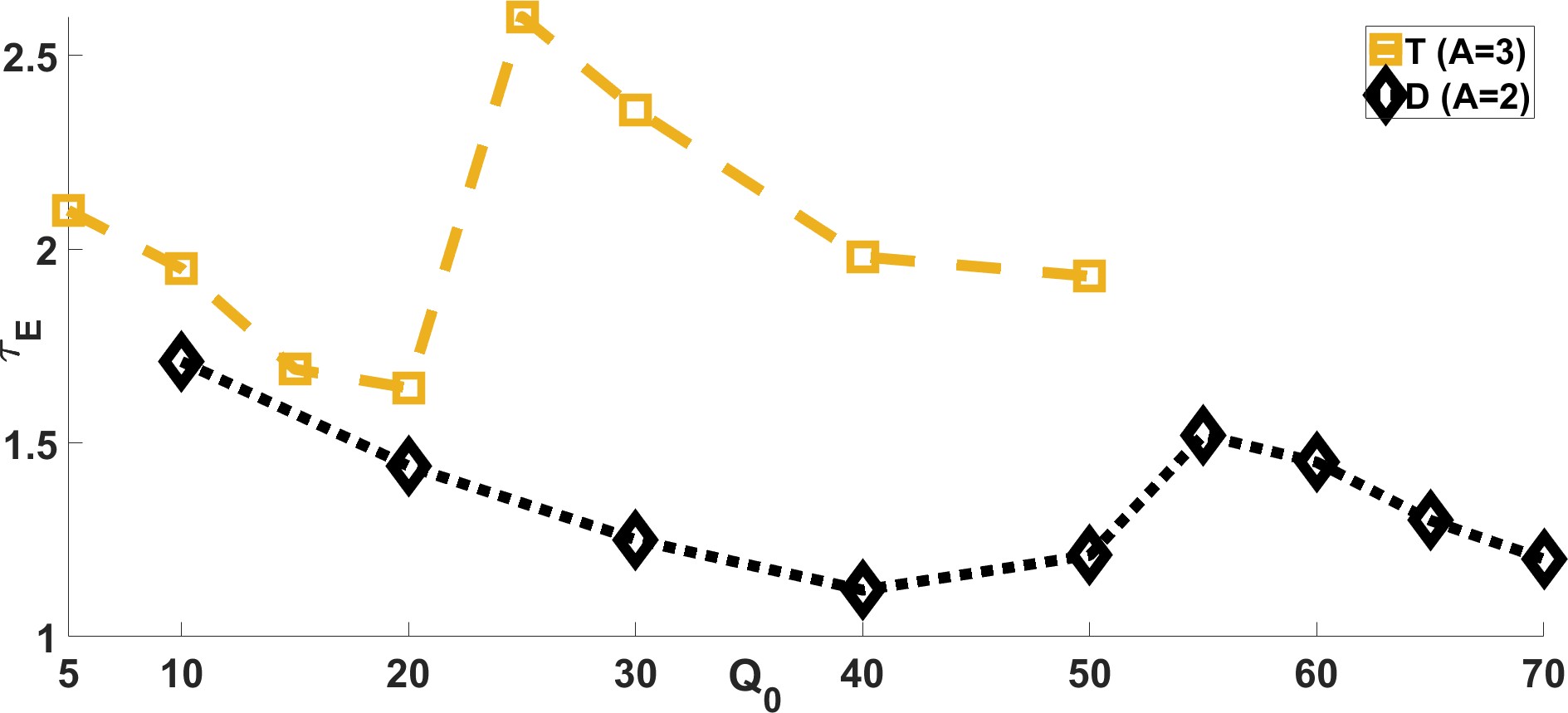}
\caption{\label{fig:Fig6} Radially averaged confinement time over the simulation domain, $\tau_E$ in ms, with respect to the radially averaged heat flux $Q_0$ in arbitrary units. For Deuterium (as shown in Fig.~\ref{fig:Fig1}) in black dotted line, for Tritium in yellow dashed line.}
\end{figure}

The dependency of the power threshold on the ion mass is crucial for magnetic confinement fusion, hence the validity of the physics embedded in our fluid turbulence model is further explored by investigating the impact of the ion mass on the power threshold. We change the ion mass from Deuterium (D) to Tritium (T) and keep constant the physical size of the box and the maximum $k_y$.  Since lengths are  normalized to $\rho_i\propto \sqrt{A}$, the size of the simulated radial box has been reduced to 152 points in the T case versus 190 in the D case. To keep $k_{y,max}$ constant, the number of poloidal modes considered in the T case has been reduced from 250 to 180. Other parameters are kept constant. The input power is scanned in the T case, similarly to what was done for the D case. The results are displayed on Fig.~\ref{fig:Fig6}.\\

\begin{figure}
\includegraphics[width=1.0\columnwidth]{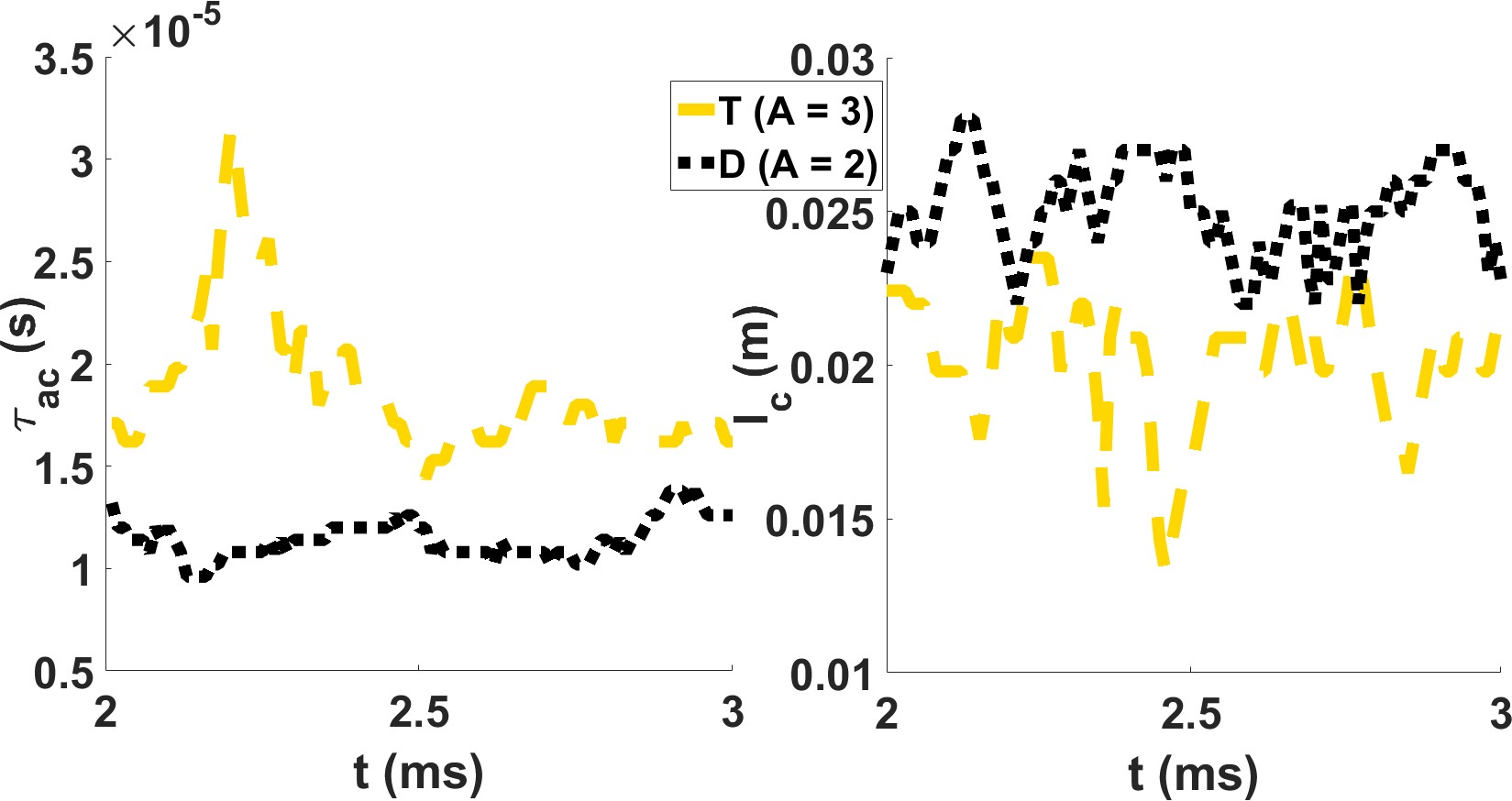}
\caption{\label{fig:Fig7} For a L mode heat source: $Q_0=10$. Left: turbulence auto-correlation time $\tau_{ac}$ for the D
case (dotted black line) and the T case (dashed yellow line). Right: turbulence auto-correlation length, $l_{c}$ for the D case (dotted black line)
and the T case (dashed yellow line).}
\end{figure}

The modelled power threshold for the pedestal formation is found to scale favorably with the ion mass: $P_{LH} \propto 1/A^{\alpha}$  with $\alpha\simeq 1.8\pm0.6$. Interestingly, at a given source, for example at $Q_0=10$, the confinement time is larger in T than in D. The auto-correlation time and length (resp. $\tau_{ac}$ and $l_c$) in D and T at $Q_0=10$ are compared in Fig.~\ref{fig:Fig7}. The auto-correlation length is computed by taking the e-folding length of the auto-correlation function $C(d_r)=\frac{<\Phi(r,y,z,t)\Phi(r+dr,y,z,t)>}{<|\Phi(r,y,z,t)|^2>}$. $\tau_{ac}$ is larger in T than in D, reflecting a lower turbulence drive in T than in D. This eases the transition to H-mode in virtue of the BDT criterion. $l_c$  is higher in D than in T, hence it is not following a Bohm scaling where $l_c\propto A^0$ nor a gyro-Bohm scaling with $l_c\propto A^{1/2}$. These results agree qualitatively with L mode experiments in ASDEX-Upgrade \cite{Sch17} and in JET-ILW \cite{Del17}.\\
Summarizing, for the first time, three key ingredients of the edge turbulent transport are simultaneously included in a flux driven turbulence code applied on realistic L mode edge parameters, namely:
\begin{itemize}
\item an edge turbulence modelling accounting for resistive ballooning modes as well as Drift Waves \cite{Rog98, Sco05, Ded19, Bon19}
\item the electromagnetic effects on edge turbulence \cite{Rog98, Sco05, Ded19, Bon19}
\item the neoclassical friction with a realistic radial variation of collisionality from banana to Pfirsch-Schl{\"u}ter regimes \cite{Cho14, Cho15, Par15}
\end{itemize}
The existence of an injected power above which a pedestal forms is recovered. The pedestal formation is shown to be due to the $\vec{E}\times\vec{B}$ shear of the turbulence, following the BDT criterion. The neoclassical friction dominates over the Reynold and Maxwell stresses both before and after the pedestal is formed.\\
The validity of the physics embedded in the fluid turbulence modelled is further challenged by changing Deuterium for Tritium. A lower value for the power threshold leading to the formation of a pedestal is observed in T vs D, similarly to experimental observations. The $\vec{E}\times\vec{B}$ quenching is made easier in T due to longer turbulence auto-correlation time, $\tau_{ac}$.\\
This work has demonstrated that the three above ingredients capture the minimal physics that has to be embedded to obtain a pedestal formation. The dynamic of the transition could be studied in further works by setting a heating ramp during one simulation \cite{Cho15} rather than changing the source from simulation to simulation as done is the present work. Moreover, further investigations of other well-known parametric dependences of the L-H power threshold need to be carried out, such as the impact of the magnetic field strength, as well as of the density. The latter would require adding an equation on particle transport in EMEDGE3D, similarly to what is done in GDB \cite{Zhu17, Zhu18}. Also the role of the SOL impact on the $\vec{E}\times\vec{B}$ shear should be explored either by testing various boundary conditions in EMEDGE3D or by coupling self-consistently the SOL \cite{Ras16, Gal17, Cha17, Par18} while keeping 3D, electromagnetic and diamagnetic effects as well as a realistic neoclassical friction. 

\section*{Acknowledgements} 
The authors acknowledge fruitful discussions with the DSC team at the PIIM UMR 7345 in particular with O. Agullo, M. Muraglia and Y. Camenen, and also to the members of the GMICS group, especially to J. Morales. Additional special thanks to E. Solano and F. Parra for discussions about the isotope effect. This work has been carried out within the framework of the EUROfusion Consortium and has received funding from the European research and training program 2014-2018 and 2019-2020 under grant agreement N 633053. The views and opinions expressed herein do not necessarily reflect those of the European Commission.

\end{document}